# Hybrid Bloch-Anderson localization of light

Simon Stützer,[1] Yaroslav V. Kartashov,[2,3,*] Victor A. Vysloukh,[2] Vladimir V. Konotop,[4]
Stefan Nolte,[1] Lluis Torner,[2] Alexander Szameit[1]

[1]*Institute of Applied Physics, Friedrich-Schiller-Universität Jena, Max-Wien-Platz 1, 07743 Jena, Germany*
[2]*ICFO-Institut de Ciencies Fotoniques, and UniversitatPolitecnica de Catalunya, 08860 Castelldefels (Barcelona), Spain*
[3]*Institute of Spectroscopy, Russian Academy of Sciences, Troitsk, Moscow Region, 142190, Russia*
[4]*Centro de Física Teórica e Computacional and Departamento de Física, Faculdade de Ciências Universidade de Lisboa,
Lisboa 1649-003, Avenida Professor Gama Pinto 2, 1649-003 Lisboa, Portugal*

We investigate the interplay of Bloch oscillations and Anderson localization in optics. Gradual washing out of Bloch oscillations and the formation of nearly-stationary averaged intensity distributions, which are symmetric for narrow and strongly asymmetric for broad input excitations, are observed experimentally in laser-written waveguide arrays. At large disorder levels Bloch oscillations are completely destroyed and both narrow and wide excitations lead to symmetric stationary averaged intensity distributions with exponentially decaying tails.

Today, there are two main localization mechanisms known in optics: Bloch oscillations [1-4] and Anderson localization [5-7]. Both rely on the transformation of infinitely extended eigenstates into spatially localized states, but the underlying mechanisms are substantially different. For an ordered system with an external linear potential gradient one finds an equidistant Bloch eigenvalue spectrum with localized eigen-states [8]. In this case, the shape of an arbitrary input wave packet is periodically restored after some propagation distance that is dictated by the potential gradient. In contrast, for Anderson localization in a disordered system the spectrum is irregular with eigenvalues corresponding to exponentially localized states, whose excitation by the input wave packet results in irregular beatings.

Naturally, it is interesting to analyze the interplay of disorder and periodicity in the presence of external potential gradient [9]. It was proven that a bounded potential in the presence of gradient does not have bound square-integrable states in continuous model, whereas in a tight-binding model the opposite happens [10]. Previous papers predicted a slow dephasing of Bloch oscillations in semiconductor super-lattices due to weak disorder [11], addressed the impact of point defects [12] and longitudinal refractive index modulations [13,14] on Bloch oscillations and localization dynamics. The problem was considered in linear [15] and nonlinear [16] two-dimensional tight-binding models in the context of damped Bloch oscillations of Bose-Einstein condensates [17]. Nevertheless, the destruction of Bloch oscillations and gradual transition to Anderson localization were never observed directly in optical settings.

In this Letter we address the interplay between Bloch oscillations and Anderson localization in an optical setting. We consider the propagation of a light in a waveguide array with a linear refractive index gradient described by a Schrödinger equation for the light field amplitude $\psi$:

$$i\frac{\partial \psi}{\partial z} = -\frac{1}{2}\frac{\partial^2 \psi}{\partial x^2} - R(x)\psi - \alpha x \psi, \quad (1)$$

where $x, z$ are the transverse and longitudinal coordinates, respectively; $R(x) = \sum_j n_j \exp[-(x-jd)^6 / w^6]$ describes the refractive index profile in the array of waveguides with width $w$, period $d$ and refractive index $n_j$; $\alpha$ is the refractive index gradient. The refractive index of each guide fluctuates within the interval $[n_{\mathrm{av}} - n_{\mathrm{d}}, n_{\mathrm{av}} + n_{\mathrm{d}}]$, where $n_{\mathrm{av}}$ is the average refractive index and $n_{\mathrm{d}}$ sets the degree of disorder. In accordance with our experiments we set $w = 0.3$ (3 $\mu$m-wide waveguides), $d = 1.3$ (13 $\mu$m period), and an averaged refractive index $n_{\mathrm{av}} = 7.2$ (real refractive index contrast $\delta n_{\mathrm{exp}} \sim 5 \times 10^{-4}$ at $\lambda = 633$ nm). 100 mm long samples correspond to a propagation distance of $z = 70$. In experiments the refractive index gradient was fixed as $\alpha = 0.173$, but in simulations we varied both $\alpha$ and $n_{\mathrm{d}}$.

In our simulations we generated $N = 10^3$ realizations of disordered arrays for each $\alpha, n_{\mathrm{d}}$ and solved Eq. (1) for each realization of $R_l(x)$. We calculated the averaged output intensity distribution $I_{\mathrm{av}}(x,z) = N^{-1}\sum_{l=1}^{N}|\psi_l(x,z)|^2$, the integral output beam center $x_{\mathrm{av}}(z) = (NU)^{-1}\sum_{l=1}^{N}\int_{-\infty}^{\infty} x|\psi_l(x,z)|^2 \, dx$, and the variance $\sigma_{\mathrm{av}}(z) = \left[(NU)^{-1}\sum_{l=1}^{N}\int_{-\infty}^{\infty} x^2|\psi_l(x,z)|^2 \, dx - x_{\mathrm{av}}^2\right]$, where $U = \int |\psi_l(x,z)|^2 \, dx$ is the conserved total power. To minimize radiation we use the input beam $\psi|_{z=0} = A(x)\exp(-x^2/x_{\mathrm{env}}^2)$, where $A(x)$ is the profile of the Floquet-Bloch mode with momentum $k=0$ taken from the first band of the ordered array with $\alpha = 0$. The width of the Gaussian envelope was $x_{\mathrm{env}} = 0.5d$ for narrow excitations and $x_{\mathrm{env}} = 3.0d$ for broad excitations.

The dependencies of the averaged quantities $\sigma_{\mathrm{av}}, x_{\mathrm{av}}$ on the parameters $\alpha, n_{\mathrm{d}}$ for narrow and broad excitations are presented in Fig. 1 (shown for $z = 200$, when $\sigma_{\mathrm{av}}, x_{\mathrm{av}}$ reach their stationary values). In the absence of disorder any input beam launched into the array experiences Bloch oscillations when $\alpha \neq 0$, which are nearly symmetric around $x = 0$ for narrow excitations and asymmetric for broad excitations. The period and full amplitude of Bloch oscillations are given by $z_{\mathrm{B}} = 2\pi/\alpha d$ and $x_{\mathrm{B}} \approx \delta b/\alpha$, where $\delta b$ is the full width of the first band in the Floquet-Bloch spectrum of the periodic unperturbed lattice. The presence of weak disorder $n_{\mathrm{d}} \ll 1$ breaks the regular equidistant eigenvalue spectrum. As a result, the revival of the input beam at $z = z_{\mathrm{B}}$ is not perfect and the oscillations become slightly irregular. This irregularity becomes more pronounced with distance due to the accumulation of phase difference between the eigenmodes. Averaging over a large number of intensity distributions yields for narrow excitations a specific pattern, shown in Fig. 2(a). Bloch oscillations are washed out and a nearly stationary distribution $I_{av}$ forms, which is symmetric around $x = 0$. The number of oscillations visible in the $I_{av}$ distribution decreases with growing disorder $n_d$ [compare Figs. 2(a) and 2(b)]. The oscillations in $\sigma_{av}$ decay almost linearly with $z$, and at sufficiently large distance $\sigma_{av}$ approaches the asymptotic value shown in Fig. 1(a). For $n_d \to 0$ the width of $I_{av}$ distribution is determined by $x_B$, whereas with growing $n_d$ the oscillations rapidly diminish [Fig. 2(c)]. Notice that for narrow beams the width of Anderson-localized intensity distribution in flat array with $\alpha = 0$ decreases with $n_d$ and becomes comparable with the amplitude $x_B$ of Bloch oscillations in regular array with $\alpha = 0.173$ at $n_d \approx 0.65$. For narrow excitations one always finds $x_{av} \approx 0$. The impact of the refractive index gradient on $\sigma_{av}$ is shown in Fig. 1(b). For small $n_d$ the dispersion $\sigma_{av}$ is remarkably close to half of the amplitude of "pure" Bloch oscillations $x_B/2$ [compare the red curve, showing $x_B/2$, and curve 1 in Fig. 1(b)]. For strong disorder dispersion is determined by $n_d$, i.e., a transition between localization due to the Bloch oscillations [Fig. 2(b)] and Anderson localization [Fig. 2(c)] occurs. This phenomenon is also responsible for a deviation of curves 1 and 2 in Fig. 1(b) at $\alpha \to 0$, since the width of the averaged pattern at $\alpha \to 0$ is determined by $n_d$ solely. The effect of increasing refractive index gradient on averaged dynamics is illustrated in Figs. 2(d)-2(f).

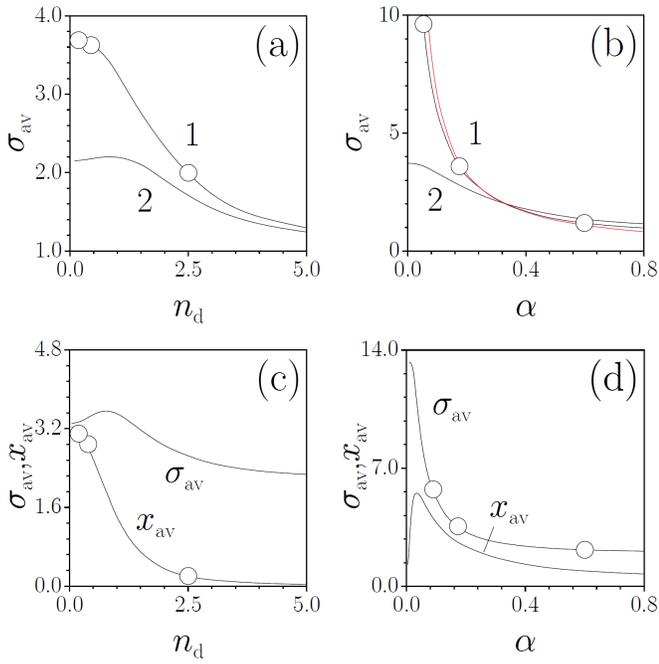

Fig. 1. Narrow excitation: (a) $\sigma_{av}$ versus $n_d$ at $\alpha = 0.173$ (curve 1) and $\alpha = 0.3$ (curve 2). (b) $\sigma_{av}$ versus $\alpha$ at $n_d = 0.5$ (curve 1) and $n_d = 1.5$ (curve 2). Red curve shows $x_B(\alpha)/2$ dependence for regular array. Broad excitation: $\sigma_{av}$ and $x_{av}$ versus $n_d$ at $\alpha = 0.173$ (c) and versus $\alpha$ at $n_d = 0.5$. The circles in (a),(b) and (c),(d) correspond to distributions in Figs. 2 and 3, respectively.

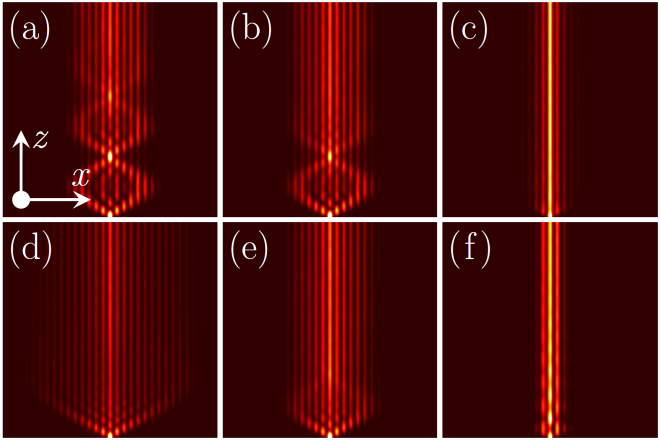

Fig. 2. Averaged intensity distributions for narrow input at $\alpha = 0.173$ for $n_d = 0.2$ (a), $n_d = 0.3$ (b), and $n_d = 2.5$ (c). Averaged intensity distributions at $n_d = 0.5$ for $\alpha = 0.09$ (d), $\alpha = 0.173$ (e), and $\alpha = 0.6$ (f). Propagation distance is $z = 100$.

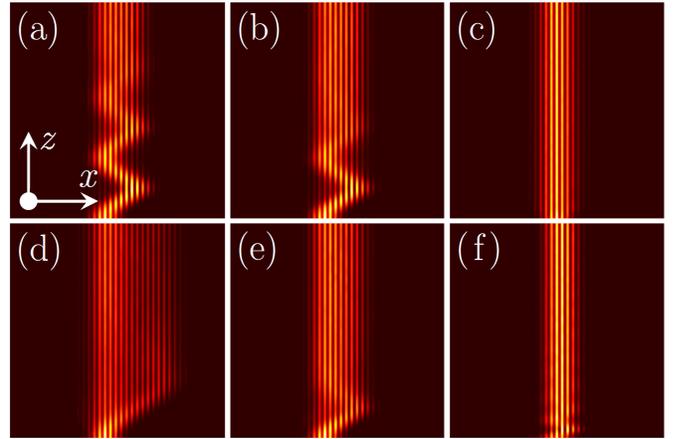

Fig. 3. The same as in Fig. 2, but for broad input excitation.

The impact of disorder is even more intriguing for broad excitations. When $n_d$ grows, one initially observes washing-out of the Bloch oscillations and a transition to a strongly localized intensity distribution [Fig. 3(a)-3(c)]. In contrast to narrow excitations where $I_{av}$ is almost symmetric, for broad excitations after large propagation distance a strongly asymmetric averaged intensity distribution forms, with its center shifted with respect to the launching position $x = 0$. The increase of disorder results in a simultaneous shrinkage of the averaged pattern and a shift of its center $x_{av}$ toward $x = 0$ [Fig. 1(c)]. The dependence $\sigma_{av}(n_d)$ is non-monotonic. While $x_{av}$ vanishes for large disorder, $\sigma_{av}$ approaches a constant value that is determined by the width of the input beam. The asymmetry of the averaged intensity distribution is most pronounced for small $n_d, \alpha$ values, which is the signature of a new regime - hybrid Bloch-Anderson localization. $\sigma_{av}$ monotonically decreases with growing $\alpha$ until it reaches a limiting value, which is determined by the width of input beam [Fig. 1(d)]. Although the input beam is at normal incidence and the fluctuations of the parameters of array follow homogeneous statistics, we observe a pronounced non-monotonic behavior of $x_{av}$ as a function of $\alpha$ [Figs. 1(d) and 3(d)-3(f)]. For small gradients the disorder dominates and the beam is localized around $x = 0$. The maximal beam center displacement is observed when the deflection due to $\alpha$ and disorder compete at similar strength. For large gradients the displacement $x_{av} \to 0$.

For all disorder levels the formation of a static averaged output profile that decays exponentially in the transverse direction was observed after a sufficiently large propagation distance. In Fig. 4, using a logarithmic scale, we show how strongly asymmetric $I_{av}$ distribution is replaced with a triangular distribution (typical for Anderson localization) when the disorder is increased.

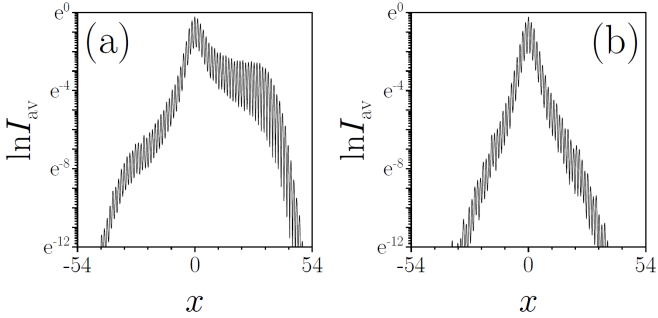

Fig. 4. Averaged output intensity distributions at $z=200$, $\alpha=0.04$ for $n_d=0.5$ (a) and $n_d=1.5$ (b).

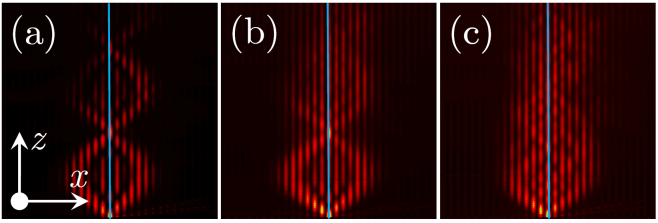

Fig. 5. Experimentally observed averaged intensity distributions in 70 mm sample showing hybrid Anderson-Bloch localization. (a) $v_d=0$ mm/min, (b) $v_d=14$ mm/min, (c) $v_d=28$ mm/min. Blue lines indicate input beam center.

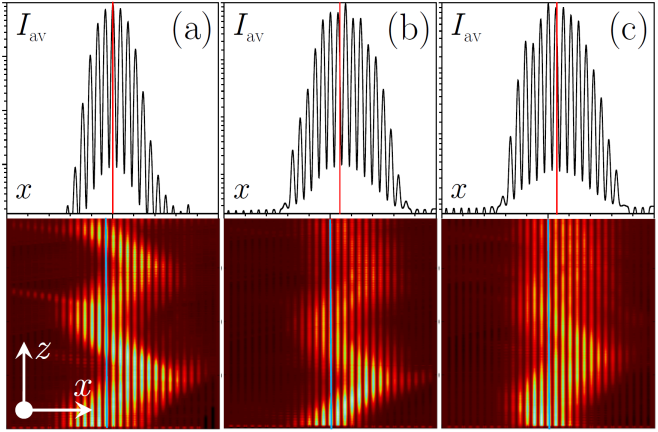

Fig. 6. Wave packet evolution for broad input in a 75 mm sample. Top row: cross section of the intensity distribution in ln scale; red lines indicate the position of integral center of the output beam.

For our experiments, we use waveguide arrays fabricated using the femtosecond laser-writing technology. Our samples, consisting of 29 waveguides each, are $L=100$ mm long. To realize a transverse linear refractive index gradient in the array, we slightly curve the array [2,18] with a curvature radius of $R=1120$ mm, yielding a Bloch period of $z_B \sim 38$ mm. The disorder is introduced by choosing randomly the writing velocity from the uniform distribution $[v_{av}-v_d, v_{av}+v_d]$, with $v_{av}=100$ mm/min and $v_d$ determining the disorder strength. The intensity distribution is monitored by fluorescence microscopy. The averaging was done over 20 different random realizations.

The experimental results are summarized in Figs. 5 and 6 for narrow and broad excitations, respectively. In Fig. 5(a), for narrow excitations and no disorder ($v_d=0$ mm/min), conventional Bloch oscillations are observed. When intermediate disorder is introduced at $v_d=14$ mm/min ($n_d\approx 0.3$) the Bloch oscillations are washed out, and the initially dynamic pattern approaches a static symmetric distribution centered at the excited waveguide, that decays almost exponentially [Fig. 5(b)]. For stronger disorder at $v_d=28$ mm/min ($n_d\approx 0.5$) the transition from Bloch oscillations to Anderson localization occurs faster and static distribution becomes narrower [Fig. 5(c)]. For broad beams at $v_d=0$ mm/min usual Bloch oscillations occur [Fig. 6(a)]. In this case the output intensity distribution is symmetric, but shifted due to the oscillations. At $v_d=14$ mm/min Bloch oscillations are partially washed out and the dynamic intensity distribution approaches a static one, which is shifted and asymmetric - a signature of hybrid localization [Fig. 6(b)]. For strong disorder with $v_d=28$ mm/min asymmetric static pattern forms faster [Fig. 6(c)].

Summarizing, we observed a gradual transition between Bloch oscillations and Anderson localization. We identified a new localization regime for broad excitations, where disorder results in the formation of asymmetric averaged intensity distributions.

### References

1. U. Peschel, T. Pertsch, and F. Lederer, Opt. Lett. **23**, 1701 (1998).
2. G. Lenz, I. Talanina, and C. M. de Sterke, Phys. Rev. Lett. **83**, 963 (1999).
3. T. Pertsch, P. Dannberg, W. Elflein, A. Bräuer, and F. Lederer, Phys. Rev. Lett. **83**, 4752 (1999).
4. R. Morandotti, U. Peschel, J. S. Aitchison, H. S. Eisenberg, and Y. Silberberg, Phys. Rev. Lett. **83**, 4756 (1999).
5. D. S. Wiersma, P. Bartolini, A. Lagendijk, and R. Righini, Nature **390**, 671 (1997).
6. T. Schwartz, G. Bartal, S. Fishman, and M. Segev, Nature **446**, 52 (2007).
7. Y. Lahini, A. Avidan, F. Pozzi, M. Sorel, R. Morandotti, D. N. Christodoulides, and Y. Silberberg, Phys. Rev. Lett. **100**, 013906 (2008).
8. G. H. Wannier, Phys. Rev. **117**, 432 (1960).
9. F. Bentosela, V. Grecchi and F. Zironi, Phys. Rev. B **31**, 6909 (1985).
10. F. Bentosela, R. Carmona, P. Duclos, B. Simon, B. Souillard, and R. Weder, Comm. Math. Phys. **88**, 387 (1983).
11. E. Diez, F. Domínguez-Adame, and A. Sánchez, Microelectronic Engineering **43-44** 117 (1998).
12. S. Longhi, Phys. Rev. B **81**, 195118 (2010).
13. W. Zhang and S. E. Ulloa, Phys. Rev. B **74**, 115304 (2006).
14. R. El-Ganainy, M.-A. Miri, and D. Christodoulides, EPL **99**, 64004 (2012).
15. A. R. Kolovsky, Phys. Rev. Lett. **101**, 190602 (2008).
16. T. Schulte, S. Drenkelforth, G. Kleine Büning, W. Ertmer, J. Arlt, M. Lewenstein, and L. Santos, Phys. Rev. A **77**, 023610 (2008).
17. S. Drenkelforth, G. Kleine Büning, J. Will, T. Schulte, N. Murray, W. Ertmer, L. Santos, and J. J. Arlt, New J. Phys. **10**, 045027 (2008).
18. N. Chiodo, G. Della Valle, R. Osellame, S. Longhi, G. Cerullo, R. Ramponi, P. Laporta, and U. Morgner, Opt. Lett. **31**, 1651 (2006).